\documentclass[useAMS, a4, usenatbib]{mn2e}

\usepackage{epsfig}
\usepackage{floatflt, epsfig}
\usepackage{epstopdf}

\def \chisq  {\ifmmode  \chi^2   \else  $\chi^2$  \fi}  
\def \spose#1{\hbox  to 0pt{#1\hss}}  
\def \lta{\mathrel{\spose{\lower 3pt\hbox{$\sim$}}\raise  2.0pt\hbox{$<$}}}
\def \gta{\mathrel{\spose{\lower  3pt\hbox{$\sim$}}\raise 2.0pt\hbox{$>$}}}

\def \kms {\ifmmode  \,\rm km\,s^{-1} \else $\,\rm km\,s^{-1}  $ \fi }
\def \kpc {\ifmmode  {\rm~kpc}  \else ${\rm~kpc}$\fi}  
\def \pc {\ifmmode  {\rm~pc}  \else ${\rm~pc}$ \fi  }  
\def \Gyr {\ifmmode  {\rm~Gyr}  \else ${\rm~Gyr}$\fi}
\def \Msun {\ifmmode \mathrm{M}_{\odot} \else $\mathrm{M}_{\odot}$\fi} 
\def \Lsun {\ifmmode L_{\odot} \else $L_{\odot}$ \fi} 
\def \Rsun {\ifmmode R_{\odot} \else $R_{\odot}$ \fi} 
\def \Msunpyr {\ifmmode M_{\odot}{\rm~yr}^{-1} \else $M_{\odot}{\rm~yr}^{-1}$ \fi} 
\def \hMsun {\ifmmode h^{-1}\,\rm M_{\odot} \else $h^{-1}\,\rm M_{\odot}$ \fi}

\def \LCDM {\ifmmode \Lambda{\rm CDM} \else $\Lambda{\rm CDM}$ \fi}
\def \sig8 {\ifmmode \sigma_8 \else $\sigma_8$ \fi} 
\def \OmegaM {\ifmmode \Omega_{\rm M} \else $\Omega_{\rm M}$ \fi} 
\def \OmegaL {\ifmmode \Omega_{\rm \Lambda} \else $\Omega_{\rm \Lambda}$\fi} 
\def \Omegab {\ifmmode \Omega_{\rm b} \else $\Omega_{\rm b}$ \fi}
\def \Deltavir {\ifmmode \Delta_{\rm vir} \else $\Delta_{\rm vir}$ \fi}
\def \rhocrit {\ifmmode \rho_{\rm crit} \else $\rho_{\rm crit}$ \fi}
\def \rhou {\ifmmode \rho_{\rm u} \else $\rho_{\rm u}$ \fi}
\def \zc {\ifmmode z_{\rm c} \else $z_{\rm c}$ \fi}

\def \rhos {\ifmmode \rho_{\rm s} \else $\rho_{\rm s}$ \fi} 
\def \rs {\ifmmode r_{\rm s} \else $r_{\rm s}$ \fi} 
\def \cvir {\ifmmode c_{\rm vir} \else $c_{\rm vir}$ \fi} 
\def \Rvir {\ifmmode r_{\rm vir} \else $R_{\rm vir}$ \fi}
\def \Vvir {\ifmmode V_{\rm  vir} \else  $V_{\rm vir}$  \fi} 
\def \Mvir {\ifmmode M_{\rm  vir} \else $M_{\rm  vir}$ \fi}  
\def \Nvir {\ifmmode N_{\rm  vir} \else $N_{\rm  vir}$ \fi}  
\def \Jvir {\ifmmode J_{\rm vir} \else $J_{\rm vir}$ \fi} 
\def \Evir {\ifmmode E_{\rm vir} \else $E_{\rm vir}$ \fi} 
\def \vvir {\ifmmode v_{\rm vir} \else $v_{\rm vir}$ \fi} 
\def \lam {\ifmmode \lambda  \else $\lambda$ \fi} 
\def \lamp {\ifmmode \lambda^{\prime} \else $\lambda^{\prime}$  \fi} 
\def \Vmax {\ifmmode V_{\rm  max} \else  $V_{\rm max}$  \fi} 
\def \Mdm {\ifmmode M_{\rm  dm} \else $M_{\rm  dm}$ \fi}

\def \Mgas {\ifmmode M_{\rm gas} \else $M_{\rm gas}$ \fi} 
\def \Mcg {\ifmmode M_{\rm cg} \else $M_{\rm cg}$\fi} 
\def \Mhg {\ifmmode M_{\rm hg} \else $M_{\rm hg}$ \fi} 
\def \Mdisc {\ifmmode M_{\rm disc} \else $M_{\rm disc}$ \fi} 
\def \Md {\ifmmode M_{\rm d} \else $M_{\rm d}$ \fi} 
\def \Mda {\ifmmode M_{\rm d,0\%} \else $M_{\rm d,0\%}$ \fi} 
\def \Mdb {\ifmmode M_{\rm d,20\%} \else $M_{\rm d,20\%}$ \fi} 
\def \Mdc {\ifmmode M_{\rm d,40\%} \else $M_{\rm d,40\%}$ \fi} 
\def \md {\ifmmode m_{\rm d} \else $m_{\rm d}$ \fi} 
\def \Mb {\ifmmode M_{\rm b} \else $M_{\rm b}$ \fi} 
\def \Mbh {\ifmmode M_{\rm b,pri} \else $M_{\rm b,pri}$ \fi} 
\def \Mbs {\ifmmode M_{\rm b,sat} \else $M_{\rm b,sat}$ \fi} 
\def \zo {\ifmmode z_{0} \else $z_{0}$ \fi} 
\def \rd {\ifmmode r_{\rm d} \else $r_{\rm d}$ \fi}
\def \rg {\ifmmode r_{\rm g} \else $r_{\rm g}$ \fi}
\def \rb {\ifmmode r_{\rm b} \else $r_{\rm b}$\fi}
\def \rs {\ifmmode r_{\rm s} \else $r_{\rm s}$\fi}
\def \rc {\ifmmode r_{\rm c} \else $r_{\rm c}$\fi}
\def \rvir {\ifmmode r_{\rm vir} \else $r_{\rm vir}$\fi}
\def \rbh {\ifmmode r_{\rm b,pri} \else $r_{\rm b,pri}$ \fi} 
\def \rbs {\ifmmode r_{\rm b,sat} \else $r_{\rm b,sat}$ \fi}

\title[The MaGICC Volume] 
{The MaGICC volume: reproducing statistical properties of high redshift galaxies}

\author[Kannan et al.] {Rahul  Kannan$^{1}$\thanks{Member of the International Max Planck Research School for Astronomy and Cosmic Physics at the University of Heidelberg (IMPRS-HD) and the Heidelberg Graduate School of Fundamental Physics (HGSFP)}\thanks{kannan@mpia.de}, Greg S. Stinson$^1$, Andrea V. Macci\`o$^{1}$,  Chris Brook$^2$, 
\newauthor{Simone M. Weinmann$^3$, James Wadsley$^{4}$, Hugh M. P. Couchman$^4$} \\ 
  $^1$ Max-Planck-Institut f\"ur Astronomie, K\"onigstuhl 17, 69117 Heidelberg, Germany \\ 
    $^{2}$ Departamento de F\'{i}s\'{i}ca Te\'{o}rica, Universidad Aut\'{o}noma de Madrid, E-28049 Cantoblanco, Madrid, Spain\\
    $^3$ Leiden Observatory, Leiden University. P.O. Box 9513, 2300 RA Leiden, The Netherlands\\ 
$^{4}$Department of Physics and Astronomy, McMaster University, Hamilton, Ontario, L8S 4M1, Canada
}

\begin{document} 
              
\date{\today}
              
\pagerange{\pageref{firstpage}--\pageref{lastpage}}\pubyear{} 
 
\maketitle 

\label{firstpage}
             
\begin{abstract}
 We present a cosmological hydrodynamical simulation of a representative volume of the Universe, as part of the Making
Galaxies in a Cosmological Context (MaGICC) project. MaGICC uses a thermal implementation for supernova and early  stellar
feedback. This work tests the feedback model at lower resolution across a range of galaxy masses, morphologies and merger
histories.
The simulated sample compares well with observations of high 
redshift galaxies ($z \ge 2$) including the stellar mass--halo mass
($M_\star-M_h$) relation, the Galaxy Stellar Mass Function (GSMF) at low masses ($M_\star <  5 \times 10^{10} \Msun $) and the number density evolution of low mass galaxies. The poor match of $M_\star-M_h$
and the GSMF at high masses ($M_\star \ge  5 \times 10^{10} \Msun$ ) indicates supernova feedback
is insufficient to limit star formation in these haloes. At $z=0$, our model produces too many stars in massive galaxies and slightly underpredicts the stellar mass around $L^\star$ mass galaxy.
 Altogether our results suggest that early stellar feedback, in conjunction with supernovae feedback, plays a major role in regulating the properties of low mass galaxies at high redshift.
\end{abstract}


\setcounter{footnote}{1}

\section{Introduction}
\label{sec:intro}

Gravitational assembly of structure in a Lambda Cold Dark Matter ($\Lambda$CDM) Universe is well understood and mostly consistent with observations.  However, the evolution of galaxies inside dark matter haloes presents many challenges for modellers.  The baryonic physics in haloes is complicated, involving various processes such as gas cooling, star formation, radiative transfer, stellar and active galactic nucleus (AGN) feedback.  These processes are highly non-linear and modelling them accurately is the major challenge for galaxy formation theory.

Fortunately, there are now large catalogues of galactic data available  from the local universe to as far back as $z=4$.  They make it possible to compare galaxy formation models with observations, throughout their evolution.  These catalogues include observations that present full spectral energy distributions of the galaxies from 1.4 Ghz radio continuum observations with the \emph{VLA} \citep{2011ApJ...730...61K}, to infrared imaging with \emph{Hubble/WFC3}, \emph{Spitzer/MIPS} and \emph{VLT/HAWK-I} \citep{2010ApJ...723..129K,2012A&A...538A..33S}. These observations give a complete picture of star formation even when it is dust obscured.

From these observations, one can construct a cosmic star formation history to compare with models \citep{1996ApJ...460L...1L,1996MNRAS.283.1388M,2004ApJ...615..209H,2008MNRAS.385..687W,2012arXiv1211.2230B}.  The shape of the cosmic star formation history has a steep rise from $z=0$ to $z=1$ before flattening off and then steadily decreasing from $z=2$ to higher redshift.

It is also possible to compare the star formation rate (SFR) of individual galaxies with their stellar mass ($M_\star$) determined from infrared photometry.  In observations, SFR and $M_\star$ show a tight correlation that is sometimes called the star forming main sequence (\citealt{2004MNRAS.351.1151B};  \citealt{2007ApJ...660L..43N}; \citealt{2011ApJ...738..106W}).  The slope of the relationship does not evolve significantly with redshift, but the normalization increases at higher redshifts (\citealt{2011ApJ...735...86W}; \citealt{2010ApJ...723..129K}).

Dividing the SFR by $M_\star$ gives the specific star formation rate (sSFR), which provides a test of the star formation efficiency compared to prior star formation.  Similar to the rise of the star forming main sequence, the sSFR rises with redshift (\citealt{2011ApJ...730...61K}; \citealt{2010ApJ...723..129K}).  Above $z=2$, some observations show that the evolution of the sSFR flattens, though \citet{2013ApJ...763..129S} found that when corrected for nebular line emission, the sSFR continues increasing up to $z=7$. 

The primary constraint used for many models is the number density of galaxies as a function of their stellar mass, the galaxy stellar mass function (hereafter, GSMF).  The GSMF is a Schecter type function characterized by a power law at low masses and an exponential cutoff.  At $z=0$, the exponential cutoff is at $M_\star\sim5\times10^{10}$ M$_\odot$ \citep{2009MNRAS.398.2177L}.  The GSMF evolves as a function of redshift: \citet{2012A&A...538A..33S} find that the low mass slope increases with redshift while \citet{2010ApJ...721..193P} finds that the slope remains constant, but the normalization increases. 

Three types of models are commonly used to understand how stars populate galaxies:
\begin{itemize}
\item Statistical models: compare statistics of simulations with observations
\item Semi-analytic models: populate dark matter haloes with stars based on halo mass, merger history, and single zone physics
\item Cosmological simulations:  Model a volume of the Universe with hydrodynamics
\end{itemize}

\subsection{Statistical Models}
The statistical models are based on comparing the GSMF with the dark matter halo mass function and lead to an understanding of how efficiently stars form as a function of dark matter halo mass.  A set of cosmological parameters makes explicit predictions about the mass function (\citealt{1974ApJ...187..425P}; \citealt{2001MNRAS.323....1S}; \citealt{2005MNRAS.359.1537R}) of dark matter haloes and how those haloes are distributed throughout the Universe.  The observed GSMF has a different shape than the dark matter halo mass function found in simulations.  The GSMF low mass slope ($\alpha$) is shallower than the low mass dark matter mass function slope.  The $M_\star\sim5\times10^{10}$ M$_\odot$ cutoff is at a lower mass than the dark matter mass function exponential cutoff. 

Halo Occupation Models make the reasonable assumption that the distribution of galaxies in the Universe is similar to the distribution of dark matter haloes, modulo some bias (\citealt{2000MNRAS.318.1144P}).  Halo Occupation Models attempt to match the correlation function statistics of galaxies and dark matter haloes to determine the stellar mass of galaxies that are most likely to be present in a particular dark matter halo.  Using Halo Occupation Modelling, one can construct a Conditional Luminosity Function that can be compared with the real luminosity function (\citealt{2003MNRAS.339.1057Y}; \citealt{2007MNRAS.376..841V}).

\citet{2006ApJ...647..201C} realized that if one used satellite masses at their time of accretion, then the clustering statistics of mass-ordered dark matter halo samples matches galaxies.  This realization lead to the abundance matching technique in which galaxies are placed in dark matter haloes with the same stellar mass ranking as that of the dark matter halo mass rank (\citealt{2009ApJ...696..620C}; \citealt{2010ApJ...710..903M}; \citealt{2010MNRAS.404.1111G}; \citealt{2010ApJ...717..379B}).  Such a match leads to the \emph{stellar mass--halo mass} ($M_\star-M_h$) relationship, the key constraint for our model.  The $M_\star-M_h$ relation consists of two power laws with a steep slope at low masses and shallow slope at high masses.  Stars form most efficiently at the break mass.  The star formation efficiency drops quickly to both higher and lower masses. The characteristic mass of the break in the power law is $M_{halo}\sim10^{12} \Msun$ at $z=0$ \citep{2013MNRAS.428.3121M}.

The availability of luminosity functions at high redshifts means that we can trace the evolution of the $M_\star-M_h$ relation. Abundance matching indicates that the $M_\star-M_h$ relation evolves surprisingly little \citep{2013ApJ...762L..31B}.  The star formation efficiency evolves most significantly to higher halo masses at higher redshift from $M_h\sim10^{12}$ M$_\odot$ at $z=0$ to $M_h\sim10^{12.5}$ M$_\odot$ at $z=3$ (\citealt{2013MNRAS.428.3121M}; \citealt{2013ApJ...770...57B}).

The key finding of the abundance matching models is that the star formation peaks earliest in the highest mass galaxies whereas, in the lowest mass galaxies, the SFR increases monotonically with time. This is a reflection of galaxy downsizing \citep{2009MNRAS.397.1776F}. This represents a delay of star formation in low mass haloes and is the most important feature that must be reproduced in models in order to get the evolution of low mass galaxies right.

\subsection{Semi-Analytic Models}
Semi-Analytic Models (SAMs) try to match the GSMF at $z=0$ using physical prescriptions based on the mass and merger history of dark matter haloes taken from simulations (\citealt{1993MNRAS.264..201K}).  SAMs show that supernovae can limit star formation in low mass galaxies \citep{1978MNRAS.183..341W,1991ApJ...379...52W,1999MNRAS.310.1087S,2003ApJ...599...38B} and that active galactic nuclei (AGN) can limit star formation in high mass galaxies (e.g. \citealt{2006MNRAS.366..499D}; \citealt{2006MNRAS.370..645B}).  

While SAMs do well matching the evolution of the high mass luminosity function, they do not match the evolution of low mass galaxies at high redshift (\citealt{2011MNRAS.413..101G}).  Current SAMs include strong stellar feedback to reproduce the GSMF at $z=0$ (e.g. \citealt{2011MNRAS.413..101G}; \citealt{2012MNRAS.422.2816B}), but the low and intermediate mass galaxies build their stellar mass at early times ($z>2$) following the assembly of the dark matter mass, because the feedback mechnism in these SAMs do not delay star formation in low mass haloes. That means that there is little evolution in the SAM luminosity functions after $z=2$, in contrast with observations (e.g. \citealt{2009MNRAS.397.1776F}; \citealt{2009ApJ...701.1765M}; \citealt{2011MNRAS.413..101G}).  The early star formation means the SAM galaxies have low specific star formation rates at $z<2$ (e.g. \citealt{2007ApJ...670..156D}; \citealt{2009ApJ...705..617D}) and high values at $z>3$ (e.g. \citealt{2010ApJ...718.1001B};\citealt{2010MNRAS.405.1690D}; \citealt{2011MNRAS.417.2737W}) although this picture might change at high redshifts due to refinement in the observational estimates of sSFR (\citealt{2013ApJ...763..129S}). This discrepancy has been looked at in detail by  \citet{2012MNRAS.426.2797W}, who use the number density evolution of low mass ($9.27<log(M_\star/\Msun)<9.77$) galaxies as a diagnostic to find that the observed evolution of the number density is not reproduced in any SAMs or simulations. They argue that the simple supernova feedback mechanism used in these models that gets the present day GSMF correct does not decouple star formation from the parent DM halo growth. 

\subsection{Simulations}

Hydrodynamical simulations  differ from SAMs  in that they include self-consistent interaction of dark matter and baryon evolution.  Although the efficiency of computational calculations has increased, it is still not possible to resolve many important physical processes, so they must be modelled at the `sub-grid' level.  These processes include gas cooling, star formation and stellar feedback.  

Since relatively little is known about star formation and feedback, the models include free parameters, which are constrained by observations.  Star formation model parameters are constrained using local observations of the Kennicut-Schmidt gas density--star formation density relation \citep{2003MNRAS.339..289S,2006MNRAS.373.1074S,2008MNRAS.383.1210S}.  The energy feedback from stars is modelled either by adding velocity to gas, called \emph{kinetic feedback}, or adding thermal energy as \emph{thermal feedback}. These models have been constrained based on observations \citep{2003MNRAS.339..289S,2006MNRAS.373.1265O,2008MNRAS.387.1431D,2009MNRAS.399.1773C,2012MNRAS.427..379M}.  The model used in this paper instead constrains stellar feedback to match the evolution of $M_\star-M_h$ relation.

There has been a lot of research on the optimal velocity for winds driven using kinetic feedback. The original models used a fixed wind velocity (\citealt{2003MNRAS.339..289S}; \citealt{2009MNRAS.399.1773C};  \citealt{2012MNRAS.427..379M}),  however, they had difficulties reproducing the GSMF at $z=0$.  Observations of metal absorption lines in outflows show that wind velocities are not constant, but are correlated with star formation rate, $v_w \approx SFR^{0.35,}$ at $z=0$ \citep{2005ApJ...621..227M} and $v_w \approx SFR^{0.3}$ at $z \approx 1.4$ \citep{2009ApJ...692..187W}. These observations motivated using momentum conserving wind models in which mass loading depends on the mass of the host galaxy such that $\dot{M}_{wind}/\dot{M}_{\star} \propto V_{circ}^{-1} $ (\citealt{2006MNRAS.373.1265O}; \citealt{2008MNRAS.387..577O}; \citealt{2011MNRAS.416.1354D,2011MNRAS.415...11D}).  Momentum conserving winds successfully reproduce the GSMF and many other observed galaxy properties at $z=0$ (\citealt{2010MNRAS.406.2325O};\citealt{2011MNRAS.416.1354D,2011MNRAS.415...11D} ; \citealt{2013MNRAS.428.2966P}), but has similar shortcomings with the low mass end of the luminosity function as the SAMs at high redshift. \citet{2012MNRAS.426.2797W} conclude that the current models of stellar feedback (in both SAMs and simulations) are unlikely to decouple the galaxy and DM halo growth due to its fundamental dependence on host halo mass and accretion history.  An alternative to the momentum driven wind model is the energy conserving approximation for driving outflows from galaxies in which the mass loading factor scales as $\dot{M}_{wind}/\dot{M}_{\star} \propto V_{circ}^{-2} $. \citet{2013MNRAS.428.2966P} find that using this approximation of a  stronger scaling of mass loading with galaxy size results in a shallower slope of the GSMF at $z=0$. The energy driven wind model also suppresses star formation at high redshift, reducing the cosmic star formation rate density to observed levels and shifting its peak to $z\sim 2.5$. This model is also successful in reproducing the GSMF at $z=1$ and $z=2$ reasonably well.

In thermal stellar feedback, stars heat the surrounding gas particles adiabatically, which creates pressure that can push gas out of galaxies (\citealt{1997RMxAC...6..261G}; \citealt{2000ApJ...545..728T}; \citealt{2003MNRAS.340..908K}; \citealt{2006MNRAS.373.1074S}).  SNe energy can only efficiently drive outflows if the Sedov-Taylor phase of gas expansion is resolved. Such resolution is infeasible even with modern computer hardware, so two
techniques have been employed to model this sub-grid phsyics. 
\citet{2006MNRAS.373.1074S} delay cooling within the blast region that a
supernova would create. \citet{2012MNRAS.426..140D} integrate all the
supernova energy that a stellar population creates and put it in a
single gas particle.  This raises the temperature to lengthen
the cooling time enough so that the hot gas particle has a
dynamical effect.

Simulations using thermal feedback have so far focused on disk structure using high resolution zoom in simulations (\citealt{2010Natur.463..203G}; \citealt{2011MNRAS.415.1051B}; \citealt{2011MNRAS.413..659S}; \citealt{2011ApJ...742...76G}; \citealt{2011MNRAS.410.1391A}).  Some recent simulations of a handful of galaxies have indicated that adiabatic feedback produces galaxies that follow $M_\star-M_h$ below $M_{\rm halo}<10^{12} \Msun$ (\citealt{2012MNRAS.424.1275B}; \citealt{2013ApJ...766...56M}). 

In most models of stellar feedback, only feedback from supernovae is considered, but \cite{2010ApJ...709..191M} recognized the amount stars can disrupt molecular clouds before any stars explode as supernovae.  \citet{2011MNRAS.417..950H} and \citet{2013ApJ...770...25A} implemented early stellar feedback schemes that rely on IR radiation pressure and tested them on isolated galaxy simulations. \citet{2011ApJ...731...91L} and \citet{2011ApJ...738...34P} found that when they mapped the pressure in different phases of the gas in the $30$ Doradus region of the LMC, UV photoheating provides more pressure than IR radiation pressure.  

In \citet{2013MNRAS.428..129S}, we assume that photo-heating from massive stars is thermalised by the time it reaches the spatial scales resolved in cosmological simulations.  So, we inject thermal energy equal to the fraction of the bolometric luminosity emitted in the UV in the time between the formation of the star and the first supernova explosion.  This early stellar feedback limits star formation to the amount prescribed by the $M_\star-M_h$ relationship and delays star formation in an L$_\star$ galaxy, so that the galaxy follows the evolution of the $M_\star$ - $M_h$ relationship.  This is a major improvement over previous galaxy formation models, as the delayed star formation means that star formation is decoupled from DM halo mass growth. Some side-effects of using early stellar feedback include transforming DM cusps to cores in galaxies up to $L_\star$ masses \citep{2012ApJ...744L...9M} and populating the circum-galactic medium with hot metal enriched gas, matching OVI observations  \citep{2013MNRAS.428..129S}. 
 
In this paper, we explore how the early stellar feedback model, described in \citet{2013MNRAS.428..129S}, affects the global properties of galaxies on a large scale. To study this we simulate a large volume of the Universe, $114$ Mpc on a side, as part of the Making Galaxies in a Cosmological Context (MaGICC) project. This simulation tests the effectiveness of our model at low resolution across a wide range of galaxy masses, environments and merger histories. We compare the properties of the galaxies in our simulations with observed statistical properties of high redshift galaxies like the GSMF, stellar to halo mass relationship, star formation rate, and the number density evolution of low mass galaxies through cosmic time.  In Section \ref{sec:sims} we briefly outline the star formation and stellar feedback mechanisms used in our simulations, in Section \ref{sec:results} we present our results at $z\ge2$ and compare them to the current observational estimates. In Section \ref{sec:dandc} we summarize our results and discuss future challenges.

\section{Simulation Method}
\label{sec:sims}

We simulate a cosmological volume, $114$ Mpc on a side, from $z=99$ to $z=2$. It is created using WMAP7 initial conditions with ($h$, \OmegaM, \OmegaL, \Omegab, \sig8 )  $=$  (0.702, 0.2748, 0.7252, 0.0458, 0.816)
\citep{2011ApJS..192...16L, 2011ApJS..192...18K}.  The simulation includes $512^3$ dark matter and $512^3$ gas particles. The dark matter particle has mass of $3.4 \times 10^8 \Msun$
and a softening length of $\sim 3.7$ \kpc. The initial gas particle mass is $6.9 \times 10^7 \Msun$ and the initial star particle mass
is $1.3 \times 10^7 \Msun$. Gas and star particles have a softening
length of $\sim 2.17$ \kpc. { \S \ref{sec:control}
describes lower resolution simulations that were used to test the resolution
dependence of our model.}

All the simulations use the smoothed particle hydrodynamics (SPH) code \textsc{gasoline} \citep{2004NewA....9..137W}.  The smoothing length is calculated using 32 nearest neighbours. Details of the physics used in the MaGICC project are detailed in 
\citet{2013MNRAS.428..129S}.  Briefly, stars are formed from gas cooler than $T$ = 10$^4$ K,
and denser than  8.7 cm$^{-3}$ according to the Kennicutt Schmidt Law
as described in  \citet{2013MNRAS.428..129S}  with the  star formation efficiency parameter $c_\star$=0.1.  The cooling used in this paper is described in detail in \citet{2010MNRAS.407.1581S}. It was calculated using CLOUDY (version 07.02; Ferland et al. 1998) including photoionisation and heating from the Haardt $\&$ Madau (unpublished) ultraviolet (UV) background, Compton cooling, and hydrogen, helium and metal cooling from $10$ to $10^9$ K.

The star particles are massive enough to represent an entire stellar population consisting of stars with masses given by the \cite{2003PASP..115..763C} initial mass function.  20\% of these have masses greater than 8 M$_\odot$ and explode as Type II supernovae from 4 until 35 Myr after the stellar population forms according to the Padova stellar lifetimes \citep{1993A&AS...97..851A, 1993A&AS..100..647B}.  Each supernova ejects $E_{SN}=10^{51}$ ergs of purely thermal energy into the surrounding gas ($\sim1$ kpc at the resolution of our simulations).  The supernova energy would be radiated away before it had any dynamical impact because of the high density
of the star forming gas.  Thus, the supernova feedback relies on delaying the cooling
based on the sub-grid approximation of a blast wave as described in \cite{2006MNRAS.373.1074S}.

The supernovae feedback does not start until 3.5 Myr after the
first massive star forms.  However, nearby molecular clouds show evidence of being blown apart
\emph{before} any SNII exploded \citep{2011ApJ...735...66M}. \citet{2011ApJ...731...91L} and \citet{2011ApJ...738...34P} found that UV photoheating is the dominant feedback mechanism in early phases of star formation by mapping out the pressure in different phases of the gas.   In simulations in the MaGICC project, like those here, $10\%$ the UV luminosity of the stars is injected into the surrounding gas over this 3.5 Myr period without disabling the cooling, at the rate of $4.45 \times 10^{48}$ erg/Myr/M$_\odot$. \citet{2013MNRAS.428..129S} showed that this energy limits star formation to the amount prescribed by the $M_\star-M_h$ relationship at all redshifts. The current work is our attempt explore how this
star formation and feedback prescription works at lower resolutions over a wide range of
galaxy masses.

\subsection{Halo identification}

For each snapshot, we find all the virialised haloes within the
high resolution region using a Spherical Overdensity (SO) algorithm.
Candidate groups with a minimum of $N_f=100$ particles are selected
using a FoF algorithm with linking length $\phi = 0.2d \approx$ $ 22 \kpc $ ($d$ is the
mean inter-particle separation).  We then: (i) find the point $C$ where
the gravitational potential is a minimum; (ii) determine the radius
$\bar r$ of a sphere centred on $C$, where the density contrast is
$\Deltavir$, with respect to the critical density of the Universe.
Using all particles in the corresponding  sphere of radius $\bar r$, we iterate the above procedure until we converge onto a stable particle set.  This stable particle set is then defined as a `halo'.  A galaxy is all stars within the particle set defined as a `halo'. This does not affect the definition of stellar mass in low mass galaxies, the focus of this paper, because their substructures contain very little amount of stars. We use a constant virial density contrast $\Deltavir = 200$, in order to be consistent with \citet{2013MNRAS.428.3121M}.  We include in
the halo catalogue all the haloes with more than 100 particles (see
\citealt{2007MNRAS.378...55M, 2008MNRAS.391.1940M} for further details on our halo finding
procedure).

\section{Results}
\label{sec:results}

We compare the simulated galaxy population in a $114^3$ Mpc$^3$ volume with a set of basic properties derived from the most recent observational estimates. These include the galaxy stellar mass function (GSMF), stellar mass--halo mass ($M_\star-M_h$) relationship, cosmic star formation history (SFH), star forming main sequence, and specific star formation rates (sSFRs).  Individual galaxies have been shown to match observations well \citep{2012MNRAS.424.1275B,2013MNRAS.428..129S}, so the volume provides an opportunity to test the accuracy and effectiveness of this feedback model at low resolution and high redshift. All the observational estimates of stellar masses and SFRs, that our results have been matched to, have been corrected to a \citet{2003PASP..115..763C} IMF.  The results presented in this paper have all been presented in co-moving units where ever applicable.

\subsection{Stellar - halo mass ($M_\star-M_h$) relation }
\begin{figure*}
\begin{center}
\includegraphics[scale = 0.5]{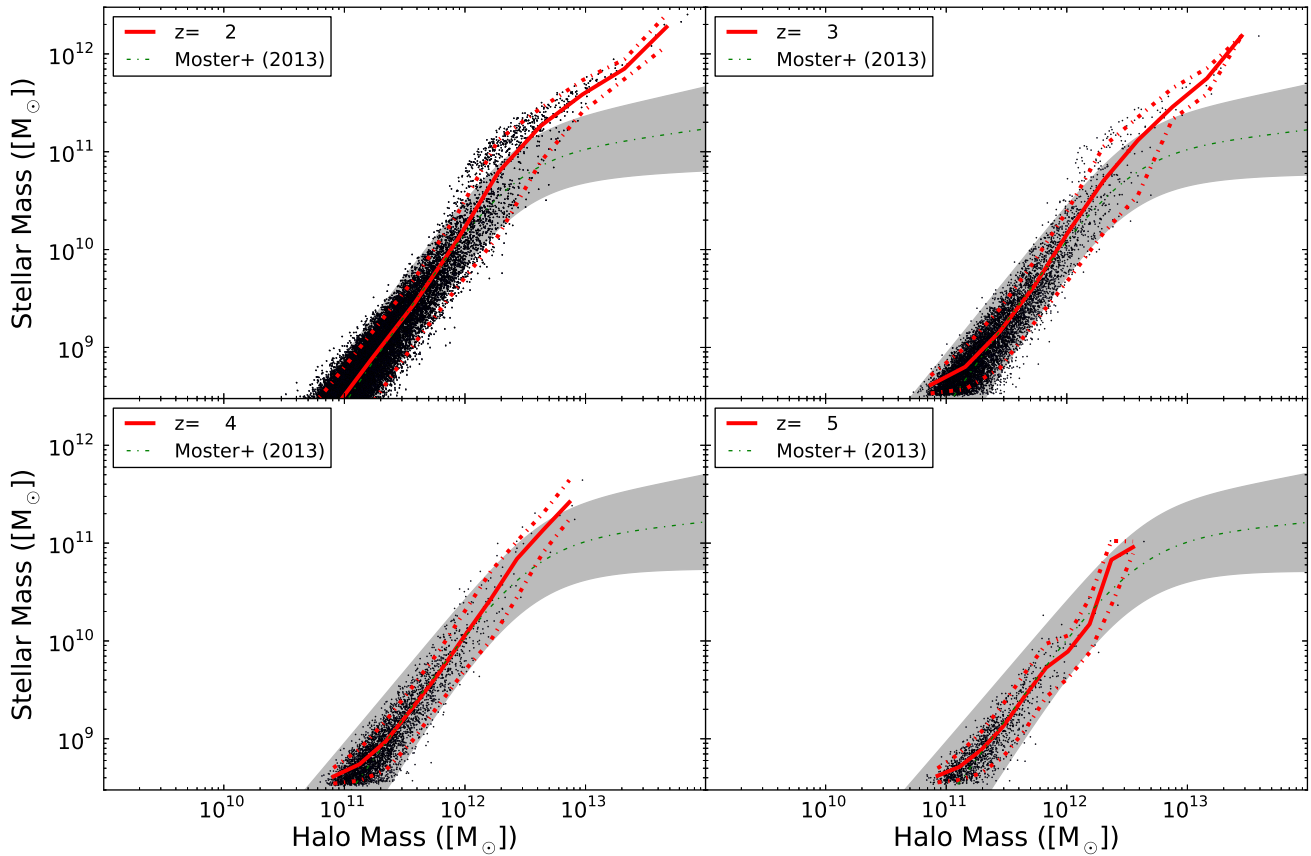}
\caption{$M_\star-M_h$ relation at different redshifts.  The black points are simulated galaxies, with the red solid line tracing the mean of the distribution and the red dotted lines indicate the $10$ and $90$ percentile limits of the distribution. The green dotted line is the \citet{2013MNRAS.428.3121M} relation derived from abundance matching techniques and the grey shaded area is the scatter derived for the relation. Our simulated galaxies match the relation  below $M_{halo} < 10^{12} \Msun$, but star formation is too efficient in high mass haloes.}
\label{sdpdm}
\end{center}
\end{figure*}

Figure \ref{sdpdm} shows the $M_\star-M_h$ relation for all the galaxies in the simulated volume (black points) that contain a minimum of $20$ star particles, or a stellar mass of  $\sim3\times10^8\Msun$.  The galaxies trace (red solid line) the slope of the $M_\star-M_h$ (green line) up to $M_{halo} = 10^{12} \Msun$ at all redshifts where it has been examined. The scatter of the simulated galaxies, quantified by the $10$ and $90$ percentile limits of the distribution (red dotted lines), also matches the variation in the relation  as obtained by \citet{2013MNRAS.428.3121M} (grey shaded area). The agreement points to the fact that the stellar feedback effectively regulates star formation to produce the right amount of stellar mass in a given halo mass at all times. 

Above a halo mass of $10^{12}$ M$_\odot$, abundance matching (green dotted line) shows a decrease in star formation efficiency. This is not reproduced in the simulation. The star formation efficiency actually increases at $M_{halo}\sim4\times 10^{12} \Msun$, before decreasing slightly as represented by the slightly shallower slope of the simulation points. The reduced SFE is due to the reduced gas accretion because of the high virial gas temperature of the halo. However, this slight decrease in SFE does not reduce the star formation in these high mass haloes to the extent observed. The implemented stellar feedback model is insufficient in these high mass haloes. Some other quenching mechanism is required such as feedback from a central super-massive black hole (AGN feedback, e.g. \citealt{2011MNRAS.410...53F}; \citealt{2005MNRAS.361..776S}).

\subsection{The galaxy stellar mass function (GSMF)}
\begin{figure*}
\begin{center}
\includegraphics[scale=0.6]{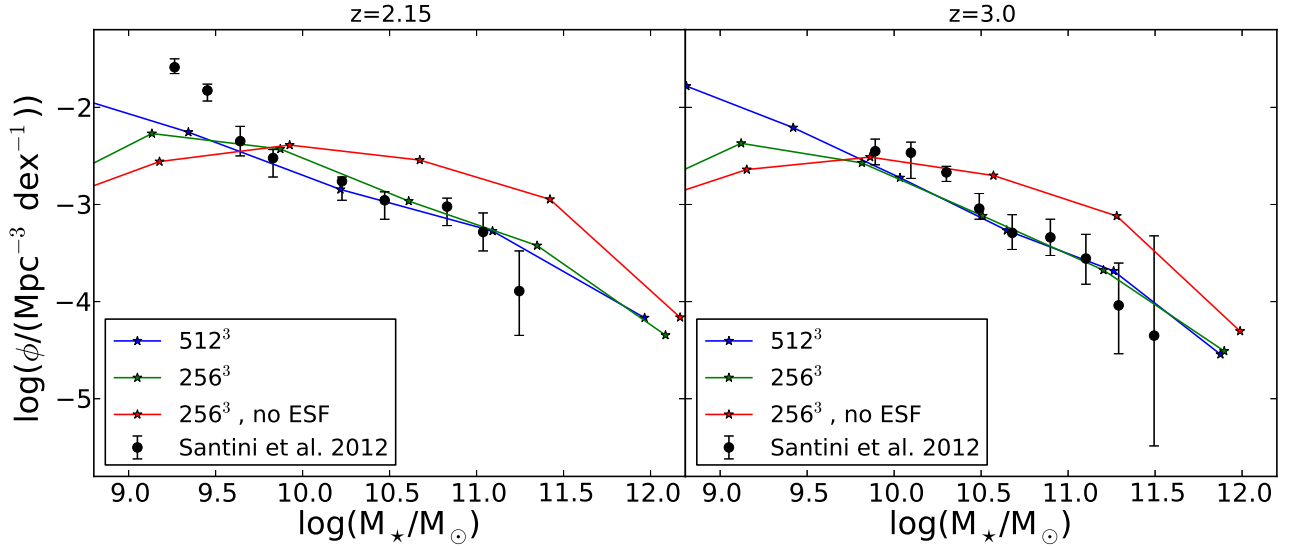}
\caption{ Galaxy stellar mass function at $z\sim2$ and $z=3$  compared to observational data taken from \citet{2012A&A...538A..33S} for three different simulations. The fiducial simulation with $512^3$ particles is shown in blue, the corresponding low resolution run ($256^3$) is shown in green, while the low resolution simulation without ESF is coloured red.}
\label{gsmf}
\end{center}
\end{figure*}

The GSMF measures the number of galaxies of a certain stellar mass in a given volume of the Universe. The era of deep, high redshift surveys has provided detailed GSMFs out to $z=3$.  We compare our simulation results to \citet{2012A&A...538A..33S}, who use deep \emph{WFC3} near-IR data complemented by  deep \emph{Hawk-I} K$_S$ band data to derive accurate stellar massees in a $\sim 33$ arcmin$^2$ area located in the {\sc goods}-South field, to study
the low-mass end of the GSMF. The observed GSMFs are presented for various redshift ranges. To compare with them, we use the simulated GSMF from the middle of the observed redshift range.  Figure \ref{gsmf} shows that the simulated galaxies from the fiducial run (blue line) trace the intermediate mass ($10^{9.5} < \rm{M}_\star/\Msun < 10^{11}$) slope of the observed GSMF (red points) very well. There is a slight discrepancy at  M$_\star<10^{9.5}\Msun$ at $z=2.15$.  This discrepancy might arise due to the the difficulty in determining the properties of low mass galaxies at such large distances or due to cosmic variance, as their data set has a small sky coverage.  The feedback model makes the slope of the GSMF as shallow as the observed value, which is non-trivial and is a major improvement over previous attempts to match the GSMF at high redshift (e.g. \citealt{2011MNRAS.413..101G}). A small discrepancy remains, as the simulated number density of high mass galaxies continues to decrease at the same rate, whereas the observations show an exponential cutoff.  This again indicates that stellar feedback is insufficient to limit star formation in these high mass haloes.  The green and red curves are control test runs, which will be discussed in \S \ref{sec:control}.

\subsection{Number density evolution of low mass galaxies}
\begin{figure}
\includegraphics[scale=0.47]{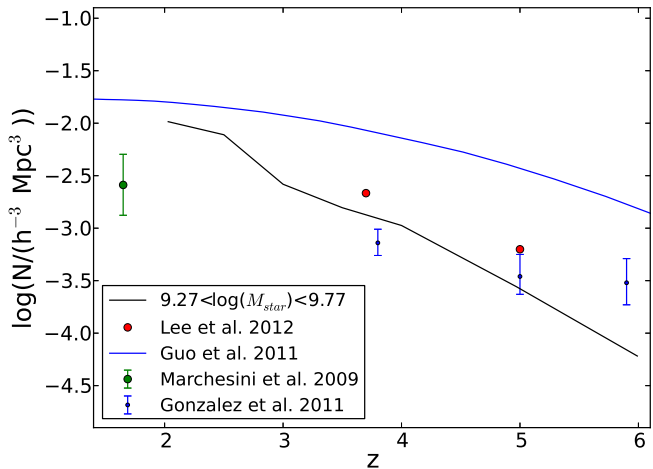}
\caption{Number density of low mass galaxies as a function of redshift (black line) compared to various observational estimates (points) and a semi analytic model (blue line) as described in \citet{2011MNRAS.413..101G}.  Our simulation results have a steeper slope and are a better fit to the observational data.}
\label{sdpdm2}
\end{figure}

\citet{2012MNRAS.426.2797W} used the number density evolution of low mass ($9.27<\mathrm{log}(M_\star/\Msun)<9.77$) galaxies to show that semi-analytic models or cosmological hydrodynamic simulations do not correctly model low mass galaxies. They argue that the simple supernova feedback mechanism changes the stellar mass at $z=0$, but renormalizes the star formation history and thus does not decouple star formation from DM accretion. \citet{2013MNRAS.428..129S} showed for a single high resolution L$^\star$ simulation that early stellar feedback can break the coupling of star formation to dark matter accretion. Figure \ref{sdpdm2} shows the number density evolution of low mass galaxies at high redshifts in our simulation volume compared to observations taken from Figure 1 of \citet{2012MNRAS.426.2797W}, as well with the SAM described in \citet{2011MNRAS.413..101G}. The simulation matches the observational results much better and lies well below the values obtained by the SAM.  The difference between the observations of \citealt{2010ApJ...713..115G} (blue points) and \citealt{2012ApJ...752...66L} (red points) is larger than the \citealt{2010ApJ...713..115G} error bars.  The simulated curve falls in the middle of these observations in contrast with the SAM that lies an order of magnitude above the observations. We note that the slope obtained from our model is still slightly steeper than observed, indicating that the simulation is building low mass galaxies faster than observed.

\subsection{Star formation History}
\begin{figure}
\begin{center}
\includegraphics[scale=0.47]{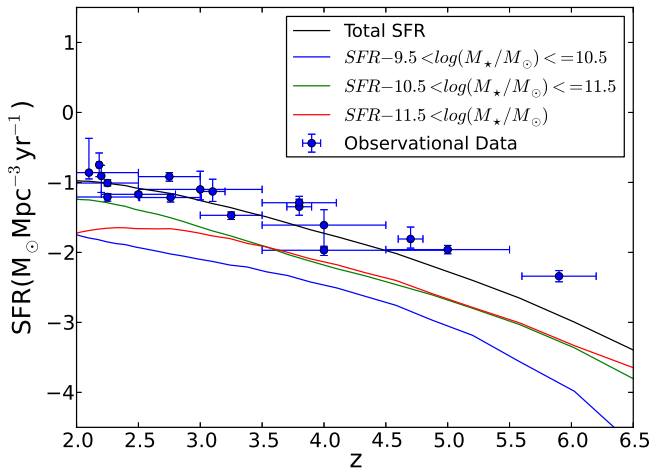}
\caption{The evolution of the star formation rate density. The blue points are a compilation of star formation rate density estimates taken from \citet{2013MNRAS.428.3121M}. The black solid line is our result for all galaxies in our volume.  The coloured curves show the star formation histories of galaxies in a certain mass range. The mass of galaxies quoted is calculated at $z=2$. The blue line is the SFR density for low mass galaxies  ($9.5<\mathrm{log}(M_\star/\Msun)<10.5$), the green for intermediate ($10.5<\mathrm{log}(M_\star/\Msun)<11.5$) mass galaxies and the red line for high mass ($\mathrm{log}(M_\star/\Msun)>11.5$) galaxies.  There is a clear trend of decreasing star formation at $z \le 3.5$ in the highest mass galaxies.}
\label{sdpdm3}
\end{center}
\end{figure}
 
We can also compare our simulation with the total number of stars formed in the Universe as a function of time. Figure \ref{sdpdm3} shows how the cosmic star formation rate evolves as a function of redshift (`Lilly-Madau plot') in our simulated volume. The observed points used for comparison are taken from \citet{2013MNRAS.428.3121M} and include star formation estimates derived from rest frame UV
(\citealt{2007ApJS..173..267S}; \citealt{2010A&A...523A..74V}; \citealt{2011MNRAS.413.2570R}; \citealt{2011ApJ...737...90B};
\citealt{2012A&A...539A..31C}), H$\alpha$ (\citealt{2011ApJ...726..109L}), combined UV and IR
(\citealt{2007ApJ...670..301Z}; \citealt{2010ApJ...723..129K}), FIR (\citealt{2010ApJ...718.1171R}) and
radio observations (\citealt{2009ApJ...690..610S}; \citealt{2009MNRAS.394....3D}; \citealt{2011ApJ...730...61K}).  The total SFR density (black line) passes through the observations from $z=2-5$.

The total SFH can be divided into separate lines based on the stellar mass of the halo at $z=2$ in which the stars are formed. The lowest mass galaxies ($9.5<\mathrm{log}(M_\star/\Msun)<10.5$) contribute little to the overall SFR density, while the the intermediate ($10.5<\mathrm{log}(M_\star/\Msun)<11.5$) and high mass ($\mathrm{log}(M_\star/\Msun)>11.5$) contribute equally up to $z=3$. Below this redshift, the SFR flattens out in the highest mass galaxies. This flattening is not sufficient to explain the quenching of high mass galaxies as shown by the failure of the simulated $M_\star-M_h$ and GSMF relations at the high mass end.  We note that our match of the star formation history is not greatly affected by the excess star formation in galaxies with ($\mathrm{log}(M_\star/\Msun) > 11.5$) because even though the galaxies in that mass range form too many stars at $z \le 3.5$, they are not the dominant population of galaxies at those redshifts.

\subsection{Star forming main sequence}
Observations show that star-forming galaxies have a tight correlation between  their SFR and $M_\star$ (e.g., \citealt{2007A&A...468...33E}; \citealt{2009ApJ...698L.116P}; \citealt{2011ApJ...738..106W}; \citealt{2012ApJ...754L..29W}). This correlation has been called the ``star forming main sequence.'' 

We compare the SFRs of our simulated galaxies with observational estimates by \citet{2010ApJ...723..129K} and \citet{2012ApJ...754L..29W}. \citet{2010ApJ...723..129K} studied SFR as a function of $M_\star$ for galaxies at $0.5 < z < 3.5$ in the GOODS-North field, using the K-selected sample from \emph{Subaru}-MOIRCS. They determined SFRs from
rest-frame, dust-corrected UV luminosity and the \emph{Spitzer}-MIPS 24 $\mu m$ flux. The depth of their data allowed them to constrain the slope of the SFR-M$_\star$ relation down to $M_\star = 10^{9.5} \Msun$ at $z \sim 3$. The median SFR as a function of stellar mass (green curve) from their sample of galaxies is plotted in top panels Fig. \ref{sdpdm4} at $z=2$ $\&$ $3$. The slope of their relation is close to unity for low mass galaxies at these high redshifts.  Our simulated galaxies match these observations well at $z=3$, but have nearly two times  less star formation at $z=2$.  This discrepancy at $z=2$ presents a challenge for all hydrodynamic simulations and SAMs \citep{2011MNRAS.417.2737W}. \citet{2008MNRAS.385..147D} suggested that an evolving stellar IMF is required to reduce the discrepancy in this relation out to $z=2$. 

\citet{2012ApJ...754L..29W} measure star formation rates using the NEWFIRM Medium-Band Survey from MIPS 24 $\mu m$ fluxes.  At $z > 2$ their detection limit is $\mathrm{log}(M_\star/\Msun) > 10.7$.  For these galaxies, they find a shallower, sub-linear, slope for their star forming main sequence, $SFR \propto M^{0.44}$, with a constant scatter of 0.34 dex.   Above their detection limit, our simulated galaxies (black points)  lie below the observations (red line) as seen in the top left panel of Fig. \ref{sdpdm4}. 

 Galaxies above a stellar mass of $10^{11}\rm{M_\star}$ show a slight reduction in star formation rate from the trend at lower masses. This reduction is likely the result of the high temperatures of the gas haloes surrounding these galaxies, which has a long cooling time, so gas accretion onto the disk is slightly reduced.  However, observations of such galaxies show a much more dramatic decrease in star formation that is not captured in these simulations.%When they limit their sample to blue galaxies, then their star forming main sequence becomes steeper.  

\subsection{Specific star formation rate evolution}
\begin{figure*}
\includegraphics[scale=0.38]{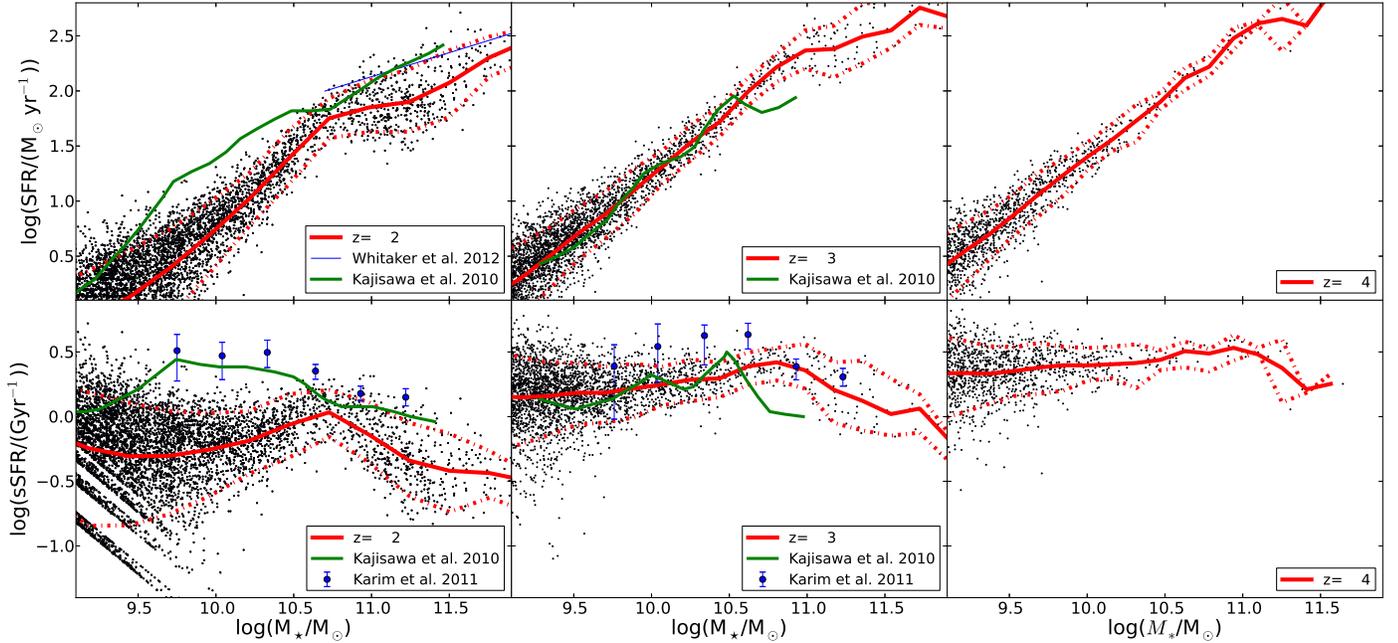}
\caption{Top panels: The star forming main sequence (black points) at different redshifts. The $z=2$ result for galaxies with $\mathrm{log}(M_\star/\Msun) > 10.7$ is matched to the observational results of \citet{2012ApJ...754L..29W} (Red line). The slope of the main sequence is much steeper at lower masses. This matches well with the observational estimates derived by \citet{2010ApJ...723..129K} at $z=2 \& 3$ (green curve).
Bottom panels: The simulated sSFR  (black points) matched to observational results matched from \citet{2011ApJ...730...61K} (Red points) and\citet{2010ApJ...723..129K} (green curve). The simulated sSFR lies below the observed values for low mass galaxies at $z=2$ by a factor of $\sim 2$, but matches very well at $z=3$.}
\label{sdpdm4}
\end{figure*}

\begin{figure}
\includegraphics[scale=0.47]{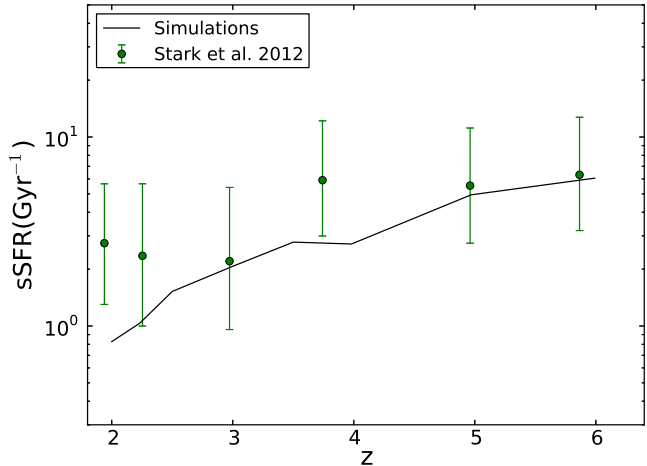}
\caption{The evolution of the specific star formation rate for sample galaxies which have stellar masses within a narrow range around $\sim 5 \times 10^9 \Msun$ (blackline). We compare the simulations with the observational estimates of \citet{2013ApJ...763..129S}. The evolution of sSFR matches well at $z\le3$, but is below the observed value at lower redshifts.}
\label{sdpdm5}
\end{figure}

Another common way to compare star formation rates with galaxy stellar masses is the specific star formation rate (sSFR), which gives the amount of star formation in haloes for a unit stellar mass of material.  As one would expect from the star forming main sequence, the bottom panels of Fig. \ref{sdpdm4} show that the simulated galaxies (black points) match the \citet{2010ApJ...723..129K} (green curve) observed sSFRs at $z=3$ but have $\sim2$ times lower sSFR at $z=2$.  We also compare our simulated results with 1.4 GHz radio continuum observations from \citet{2011ApJ...730...61K} of star formation in galaxies in the 2 deg$^2$ COSMOS field.  The simulated galaxies in our volume (black points) are in good agreement above $\mathrm{log}(M_\star/\Msun) > 10.7 $, but are $2-3$ times lower below this mass range at both $z = 2$ $\&$ $3$.  

\citet{2011ApJ...730...61K}, like other authors before them \citep{2009ApJ...697.1493S,2010ApJ...713..115G}, found that sSFR increases for galaxies in a given stellar mass range from $z=0$ to $z\sim2$, but then does not evolve much from $z=2$ to $z \sim 7$. \citet{2011MNRAS.417.2737W} shows that such observations are contradictory with most models in which higher gas accretion rates at higher redshift and lower galaxy stellar masses translate into larger sSFR in galaxies within a fixed stellar mass range. 

\citet{2013ApJ...763..129S} re-examined their data and found that their \emph{Spitzer}-IRAC photometry was contaminated by nebular emission. They use the photometric excesses in the contaminated $[3.6]$ filter to estimate the equivalent width distribution of H$\alpha$ emission at $3.8 < z < 5.0$. The corrected sSFRs increase from $z=4$ to $z=7$ by a factor of $\sim5$ similar to model predictions.  Figure \ref{sdpdm5} shows the evolution of the sSFR in simulated galaxies (black line) within a narrow stellar mass range around $\sim 5 \times 10^9 \Msun$. The simulation values are consistent with the corrected \citet{2013ApJ...763..129S} values (green points) for $z>3$.  However at $z<3$ our simulation results are below the observed relation. 

Many other authors find also find lower than observed sSFRs in their models (\citealt{2008MNRAS.385..147D}; \citealt{2012MNRAS.426.2797W}). The higher observed sSFRs again indicates delayed star formation in low mass galaxies.  Although our model does a better job of delaying the star formation at early times than most SAMs and hydrodynamic simulations, below $z=3$ the simulated haloes may still be forming too few stars. This suggests the importance of some other physical mechanism, not modelled in our simulation,  like the dependence of the star formation on gas metallicity \citep{2011ApJ...731...25K, 2012ApJ...753...16K}, that could further delay star formation at earlier times and increase the sSFR of these galaxies at $z=2$. 

\subsection{Results at $z=0$}
\begin{figure}
\includegraphics[scale=0.47]{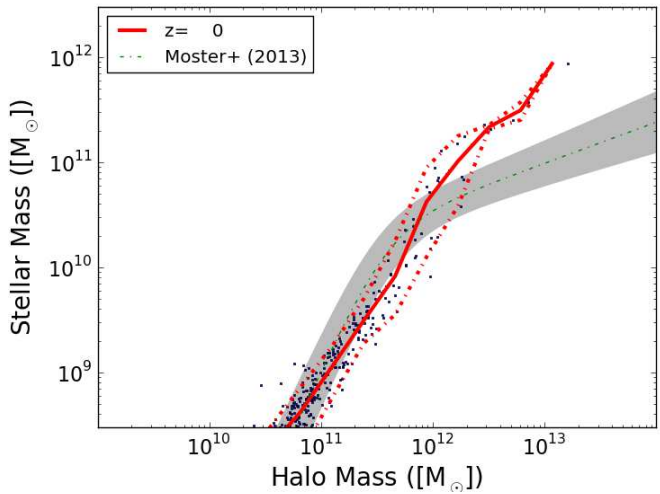}
\caption{The stellar-halo mass relation for galaxies in a $16$ $h^{-1}$ Mpc at $z=0$. Galaxies around $\rm{M}_{halo} = 10^{11} \Msun$ have about half the expected stellar mass.  Galaxies with $\rm{M}_{halo} > 10^{12} \Msun$ continue to exhibit the overcooling problem like they did at high redshift. The solid red line is the median of the simulation points, while the dotted lies show the 10 and 90 percentile limits for those bins.}
\label{sdpdm7}
\end{figure}

\begin{figure}
\includegraphics[scale=0.45]{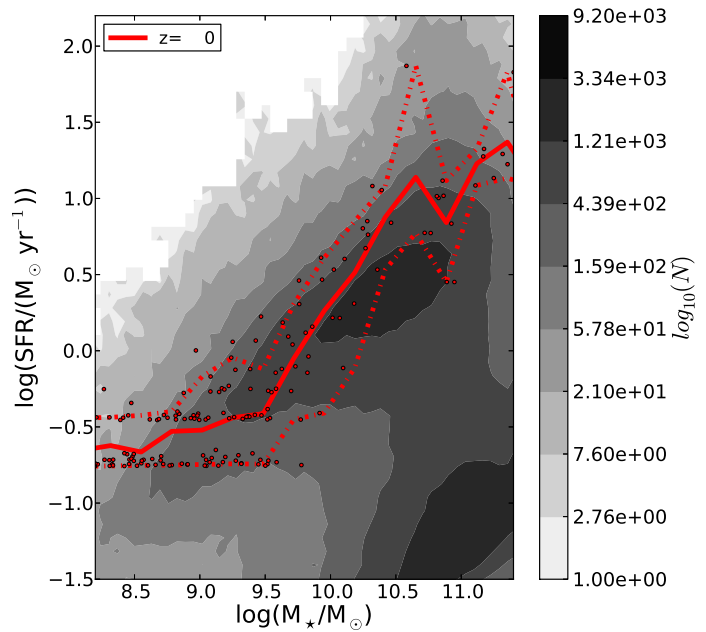}
\caption{The simulated star forming main sequence at $z=0$.  Individual galaxies
are the red points while the solid red line represents the median of those
galaxies.  The dot-dashed lines represent
the 10 and 90 percentile limits of the galaxy distribution.  The simulations are compared 
with 894k galaxies from SDSS as in \citet{2004MNRAS.351.1151B} plotted as the
grey contours.}
\label{sdpdm8}
\end{figure}

 Galaxies in the local Universe are the easiest to observe and compare with 
our model.  Unfortunately, it is too computationally demanding to simulate 
the full cosmological volume to $z=0$.   So, we select a $16$ $h^{-1}$ Mpc sub-volume from the fiducial simulation at $z=2$. The region was selected to limit the number of high mass haloes present in the region. The lack of massive haloes reduces the computational cost but also reduces the density of the region by $\sim$ $10\%$ compared to the mean density for full volume. This kind of  volume selection also impairs our ability to compare the volume weighted properties of galaxies like the GSMF. On the other hand,  the individual properties of galaxies like the stellar mass compared to the halo mass of the galaxy ($M_\star - M_h$ relation) and the star formation rate compared to their stellar mass (star formation main sequence) are expected to remain similar irrespective of the surrounding density field.  \citet{2004MNRAS.350.1385S} showed that halo formation weakly depends on the surrounding density field. So, only the $M_\star - M_h$ relation and the star formation main sequence obtained from the selected region are shown at $z=0$ in Figs. \ref{sdpdm7} and \ref{sdpdm8}.    

We include gas particles only inside the $16$ $h^{-1}$ Mpc sub-volume. Outside this region the particles are re-binned to a lower resolution in order to save computing time. The simulation was then restarted from $z=2$ and allowed to continue to $z=0$, with all the other parameters unchanged from the fiducial run. This region contained enough galaxies
for us to make a statistical comparison at $z=0$ with observations.

Fig. \ref{sdpdm7} shows the ($M_\star - M_h$) relation for the galaxies in the selected cube at $z=0$. The black points are simulated galaxies, the red solid line traces the mean of the distribution and the red dotted lines indicate the $10$ and $90$ percent confidence intervals of the distribution. The green dotted line is the \citet{2013MNRAS.428.3121M} relation derived from abundance matching techniques and the grey shaded area is the scatter derived for the relation. The simulation still provides a fair match to 
the observations at $\rm{M}_h \sim10^{11} \Msun$, though the galaxies 
at $z=0$ have half the stellar mass of the observed galaxies. 
The over-cooling problem also remains in
higher mass galaxies ($\rm{M}_h > 10^{12} \Msun$). 

 As mentioned in \S \ref{sec:intro}, previous studies using momentum driven winds SNe feedback recipes (e.g \citealt{2010MNRAS.406.2325O}, \citealt{2009MNRAS.399.1773C}) also tend to overproduce
the stellar mass of  massive galaxies and slightly under predict the stellar mass at the knee of the stellar mass function. Energy driven variable wind models seem to be capable of reproducing the low-z GSMF (for e.g. see \citealt{2013MNRAS.428.2966P}). 
Our model is more successful in reproducing the galaxy stellar
mass function at high redshift ($z>2$) in the low mass galaxy regime, while
the previous studies largely overpredict  the number of low 
mass galaxies at these high redshifts (as shown in Fig. 1 of \citealt{2012MNRAS.426.2797W}) and the differences between observations and our model at $z=0$ is pretty small and comparable to previous works. % All together this shows how our

Observations of the star forming main sequence are also more complete at $z=0$. 
Fig. \ref{sdpdm8} shows how the simulated galaxies compare to those 
observations (grey contours from \citealt{2004MNRAS.351.1151B}).  
The median star formation rates of the simulated galaxies 
are $\sim 0.5$ dex higher than the locus of the observed star forming 
main sequence, but are still within the observed range of star formation rates.
This is a change from high redshift where the simulated galaxies had
systematically lower star formation rates than observations. 

No simulated galaxies populate the quiescent--high stellar mass corner of the plot. 
Even at $z=0$, the simulation 
cannot produce red, dead galaxies most likely due to the lack of AGN feedback. 

While the simulations have trouble at high masses, this sample of galaxies
at $z=0$ suggests that the simulations model the statical properties of 
low mass galaxies well throughout the history of the Universe.

\section{Effect of resolution and early stellar feedback}
\label{sec:control}
 To test the effect of resolution and early stellar feedback,
we simulated the fiducial volume 
at a lower resolution containing 256$^3$ $2.76\times10^9$ \Msun \ dark 
matter and 256$^3$ $5.5\times10^8$ \Msun \ gas particles. 
Star particles form with masses of $1.83\times10^8$ \Msun.
The dark matter particles use a softening length of $\sim 3.7 \kpc$, 
while the gas and star particles use a softening length of $\sim 2.17 \kpc$. All the other simulation parameters are the same as used in the fiducial run.
The low resolution simulation was performed with two different feedback 
models, one with SNe feedback only, and the other adding early 
stellar feedback to the SNe feedback.   

Fig. \ref{gsmf} shows the GSMF's for these simulations (green and red lines) 
in addition to the fiducial run (blue curve). The low resolution volume 
with the same physics as the fiducial run (green curve) matches the 
fiducial run and observations for M$_\star > 10^{9.5} \Msun $.  The simulation
without early stellar feedback has too many galaxies with M$_\star > 10^{10} 
\Msun$. The decrease in the number of M$_\star > 10^{9.5} \Msun $ in the
low resolution simulations is caused by the resolution limit. These
galaxies consist of only a couple star particles, so star formation is not
well sampled and the results cannot be trusted.  Fig. \ref{gsmf} shows 
that our model is fairly well converged as well as the need for early
stellar feedback to produce realistic galaxies.

\section{Discussion and Conclusions}
\label{sec:dandc}

We examine the effect of early stellar feedback used in the Making Galaxies in a Cosmological Context (MaGICC) project on a broad sample of galaxies in a cosmological volume of $114^3$ Mpc$^3$.  The stellar feedback used is exactly the same as that used for a high resolution $L_\star$ galaxy  \citep{2013MNRAS.428..129S}.  We compare the simulated galaxies with the observed $M_\star-M_h$ relation, the galaxy stellar mass function, the cosmic star formation history, the star forming main sequence and the specific star formation rate. The simulated galaxies do a good job matching each observation to $z=2$, the time when previous models have most deviated from observations.  Our use of early stellar feedback is the key difference between our simulation and ones run previously.  The way that it delays star formation in $M_h < 10^{12} \Msun$ galaxies allows the simulations to match many observed statistical properties of high redshift galaxies.

 At $z \ge 2$, the simulated galaxies not only follow the $M_\star-M_h$ for $M_{h} < 10^{12} \Msun$ at all the redshifts examined but also match the scatter in the relation.  Correspondingly, the simulated galaxies match the shallow slope at the low mass end of the galaxy stellar mass function.  The slope of the GSMF relationship was not a constraint for the simulation, but is a natural by product of the stellar feedback recipe used.  It is a major improvement over previous attempts to match the GSMF at high redshift.  The early stellar feedback decouples the growth of stellar mass from DM mass by effectively blowing the gas away from the disc either into the circum-galactic medium or entirely out of the halo. This helps regulate the number density of low mass galaxies to the observed values by delaying star formation in these haloes.  

The simulated star formation history of the Universe also matches a variety of different observations. The model predicts that the lowest mass galaxies ($9.5<\mathrm{log}(M_\star/\Msun)<10.5$) contribute little to the overall SFR density, while the intermediate ($10.5<\mathrm{log}(M_\star/\Msun)<11.5$) and high mass ($\mathrm{log}(M_\star/\Msun)>11.5$) galaxies contribute equally up to $z=3$. After $z=3$, the star formation slows in the highest mass galaxies.  %The simulated star formation history is not greatly affected by the excess star formation in galaxies with ($\mathrm{log}(M_\star/\Msun) > 11.5$), because even though the galaxies in that mass range are forming too many stars at $z<3.5$, they are not the dominant population of galaxies at those redshifts.

At $M_{h} > 10^{12} \Msun$, too many stars form, which is shown by the presence of galaxies above the abundance matching $M_\star-M_h$ relation and the lack of an exponential cutoff in the GSMF.  These indicate that the thermal stellar feedback is unable to quench star formation like is observed in massive galaxies.  

Comparing SFR with stellar mass, the simulated galaxies lie along a tightly correlated ``star forming main sequence.''  The simulated galaxies match observations by \citet{2010ApJ...723..129K} at $z\ge 3$, but there is a slight discrepancy at $z=2$ between simulations and observations. At a given stellar mass, the simulated SFRs and correspondingly, the sSFRs, are $\sim 2$ times lower than the observed values at $9.5<\mathrm{log}(M_\star/\Msun)<10.5$.  The high sSFRs in low mass haloes at $z=2$ suggests that there needs to be a significant amount of cold gas still present in these galaxies at $z=2$. Although our model does a better job of delaying the star formation at early times than most SAMs and hydrodynamic simulations, after $z=3$ the simulated galaxies are forming too few stars. 

\citet{2008MNRAS.385..147D} showed that the higher observed SFRs at $z\le 2$ can be explained by an evolving stellar IMF, which becomes increasingly bottom-light at high redshift.  However, \citet{2009ApJ...701.1765M} showed that when such a bottom light IMF was used to model observations, the resulting observed high-redshift GSMF contained less galaxies, making the discrepancy with model GSMFs worse. 

Regarding the evolution of sSFRs at $z>3$, our simulation results are consistent with the revised \citet{2013ApJ...763..129S} observations for a sample of galaxies with stellar masses centred around $5 \times 10^9 \Msun$.  The increasing sSFR at $z \ge 4 $ is consistent with increasing baryon accretion rates at larger redshift translating into larger sSFR in galaxies of fixed stellar mass. However, our simulated galaxies have lower sSFR values than observed at $z=2$.  \citet{2012MNRAS.426.2797W} argued that the correct sSFR evolution should follow naturally from the correct evolution of the GSMF.  We see a slight deviation from the observed sSFR relation even though we match the GSMF. It must be noted that \citet{2012MNRAS.426.2797W} performed their analysis at $z<2$, while our simulation has only reached at $z=2$, where the observational estimates are less robust and and might show some internal inconsistency among different galaxy properties (e.g. sSFR and GSMF).  

There may also be another physical mechanism delaying star formation.  \citet{2011ApJ...731...25K} and \citet{2012ApJ...753...16K} argue that star formation depends sensitively on a metallicity threshold.  Until gas reaches this threshold, which coincidentally also delays the formation of H$_2$, star formation is delayed in low mass galaxies at $z>3$, which leaves sufficient cold gas at $z=2$ to increase the sSFR of these galaxies to the observed values.

 To compare the model with observations of the local Universe, the inner $16$ $h^{-1}$ Mpc of the fiducial run was simulated with gas to $z=0$. The  $M_\star - M_h$ relation is reproduced at low masses ($\rm{M}_h = 10^{11} \Msun$) and an over cooling problem still exists at high masses ($\rm{M}_h = 10^{12} \Msun$). In the intermediate mass regime, we are below the relation by a factor of two. We also match the observed star forming main sequence quite well, although we are a bit above the relation throughout the entire mass range. These results indicate that our model does not fare so well at $z=0$ as at high redshifts but the errors are low when compared to many semi-analytic models and simulations (\citealt{2011MNRAS.413..101G}; \citealt{2008MNRAS.385..147D}). 

 Two low resolution ($2 \times 256^3$ particles) realisations of the fiducial volume were simulated to test the effect of resolution and importance of ESF. Both volumes 
used the same the same physics as the fiducial volume, but one had ESF turned off.  The low resolution volume fiducial simulation compares well with 
the high resolution fiducial run and observations for galaxies with 
M$_\star > 10^{9.5} \Msun $ (20 star particles).  However, the re-simulation 
without ESF has too many galaxies with M$_\star > 10^{10} \Msun $. 

Altogether, our results suggest that stellar feedback is one of the most important factors regulating star formation in $M_{halo} < 10^{12} \Msun$ galaxies.  What is most important is \emph{when} the feedback occurs rather than simply the amount of feedback energy.  Simply increasing and decreasing the feedback energy will only set the normalisation i.e., the total stellar mass of present at $z=0$, but the key is delaying star formation in low mass galaxies.  When we include stellar feedback immediately after a star forms until supernovae stop exploding after 30 Myr, star formation is significantly delayed in low mass galaxies.  In this way, we account for the downsizing in galaxy populations by delaying the star formation in low mass galaxies with our stellar feedback model and thus reconcile a couple key aspects of a \LCDM cosmology with observations.

\section*{Acknowledgements}
 We thank the anonymous referee for insightful comments which helped us in improving the paper. We are grateful to Ben Moster for providing his data in electronic format. We thank Arjen van der Wel and Aaron Dutton for valuable discussions. The simulations were performed on the \textsc{theo} cluster of  the
Max-Planck-Institut f\"ur Astronomie at the Rechenzentrum in Garching;
the clusters hosted on \textsc{sharcnet}, part of ComputeCanada  and the Milky Way
supercomputer, funded by the Deutsche Forschungsgemein-
schaft (DFG) through Collaborative Research Center (SFB
881) ”The Milky Way System” (subproject Z2), hosted
and co-funded by the J\"ulich Supercomputing Center (JSC). We greatly appreciate
the contributions of these computing allocations.
RK, AVM and GS also acknowledge support from SFB 881 ``The Milky Way System'' (subproject A1)  of the German Research Foundation (DFG).
CBB acknowledges Max- Planck-Institut f\"ur Astronomie for its hospitality and financial support through the  Sonderforschungsbereich SFB 881 
(subproject A1) of the DFG.

\bibliographystyle{mn2e}
\bibliography{ms.bib}

\label{lastpage}

\end{document}